\documentstyle[12pt,a4,eclepsf]{article}
\def\nm{\nonumber}

\def\beqa{\begin{eqnarray}}
\def\beq{\begin{equation}}  
\def\F{{\cal{F}}}
\def\wF{\widetilde{{\cal{F}}}}
\def\G{{\cal{G}}}
\def\L{{\cal{L}}}

\def\eeqa{\end{eqnarray}}
\def\eeq{\end{equation}}
\def\lab{\label}    
\def\pa{\partial}
\def\tx{\theta_{x}}
\def\ty{\theta_{y}}

\def\l{\Lambda}

\def\wa{\widetilde{a}}

\def\sw{Seiberg-Witten }
\def\ym{Yang-Mills }
\def\pf{Picard-Fuchs }
\def\wa{\widetilde{a}}

\begin{document}

\begin{titlepage}
\thispagestyle{plain}
\pagenumbering{arabic}
\vspace*{-1.9cm}
\begin{center}
\begin{tabular}{c}
\makebox[15.5cm]{$\ $} \\ 
\hline
\end{tabular}
\end{center}
\vspace{1.0cm}
\begin{center}
{\Large \bf Non-perturbative Solutions to}
\end{center}
\vspace{-7.0mm}
\begin{center}
{\Large \bf $N=2$ Supersymmetric Yang-Mills Theories}
\end{center}
\vspace{-7.0mm}
\begin{center}
{\large \bf $-$Progress and Perspective$-$\footnote{
Note based on the talk presented in the colloquia in 
mathematics at 
RIMS, February 24 (Wed), 1999.}}
\end{center} 
\lineskip .80em
\vskip 4em
\normalsize
\begin{center}
{\large Y\H uji Ohta}
\end{center}
\vskip 1.5em
\begin{center}
{\em Research Institute for Mathematical Sciences }
\end{center}
\vspace{-8.5mm}
\begin{center}
{\em Kyoto University}
\end{center}
\vspace{-8.5mm}
\begin{center}
{\em Sakyoku, Kyoto 606, Japan.}
\end{center}
\vspace{1.0cm}
\begin{abstract}

This note reviews the progress on the low energy dynamics of 
$N=2$ supersymmetric Yang-Mills theories after the works of 
Seiberg and Witten. Specifically, the theory of 
prepotential for non-specialists is reviewed. 

\end{abstract}
\end{titlepage}

\baselineskip 6.5mm

\begin{center}
\section{Introduction}
\end{center}

\renewcommand{\theequation}{1.\arabic{equation}}\setcounter{equation}{0}

The low energy effective theory of $N=2$ supersymmetric \ym theory 
which has two kinds of supersymmetry is generated by a holomorphic 
function called prepotential and any information concerning 
the theory is available from the prepotential if it 
is determined. However, unfortunately, for this theory 
instantons are expected to contribute as a non-perturbative 
effect which can not be detected in the perturbation theory, 
and therefore the determination of the prepotential including 
this effect was not correctly proceeded so far. 

However, in the summer in 1994, Seiberg and Witten \cite{SW1,SW2} pointed out 
that in the case of SU(2) gauge theory the prepotential correctly 
including the instanton effect, 
often called non-perturbative or exact solution, can be obtained, 
provided the periods of 1-form on a Riemann surface (elliptic curve) are 
given. Their proposal was immediately extended by many authors to the 
cases of other gauge groups with or without quarks, 
and a number of papers was appeared in almost three years since then. 
Most of them was written for miscellaneous aspects of the prepotential 
and related works, but the results of these studies supported that the 
approach taken by Seiberg and Witten produced the physically 
meaningful prepotential. 

In this talk, I will show the main results concerning the prepotential 
from the works proceeded in this brief period. Specifically, I 
will talk on the method of \pf equation, 
the evaluation of the periods and the prepotentials via the Barnes 
type integral representation, 
the relation between instanton effect and the well-known 
scaling relation of prepotential of Matone, by using 
explicit examples. Finally, a perspective will be presented.     


\begin{center}
\section{$N=2$ supersymmetric \ym theory}
\end{center}

\renewcommand{\theequation}{2.\arabic{equation}}\setcounter{equation}{0}

\begin{center}
\subsection{$N=2$ supersymmetry}
\end{center}

Firstly, let us explain what is the supersymmetry \cite{WB}. 
As is well-known, there are particles called bosons with integer spins 
and fermions with half-integer spins, and these have different 
statistical properties. Supersymmetry is the symmetry which 
connects these particles. 
In the case that one boson (fermion) corresponds to one fermion (boson), 
supersymmetry is only one, so it is called $N=1$, while the case of two 
fermions (bosons) it is called $N=2$. The \ym theory to be discussed 
is the gauge theory enjoying two supersymmetries, namely, it is 
$N=2$ supersymmetric \ym theory. 

As the particles appearing in this $N=2$ \ym theory, 
there are gauge fields $A_{\mu}$ $(\mu =0,\cdots, 3)$, Weyl fermions 
$(\lambda ,\psi)$ and complex scalar field $\phi$ and 
these as a whole are referred as $N=2$ chiral multiplet or gauge 
multiplet. This chiral multiplet is often arranged by 
	\beq
	\begin{array}{ccc}
	& A_{\mu} & \\
	\lambda & & \psi \\
	& \phi & 
	\end{array}
	,\eeq
which means that $A_{\mu}$ supersymmetrically transform 
to $\lambda$ and $\psi$, etc. 
Note that in this multiplet since the gauge fields are included 
all particles belonging to this multiplet must take the value in the 
adjoint representation of the gauge group $G$ and 
the number of components (associated with $G$) is the same with 
the dimension of the gauge group.   

As an $N=2$ theory, other particles can be included, and 
in that case there is 
the multiplet called scalar or hypermultiplet consisting of 
Weyl fermions $(\psi_q ,\psi_{\widetilde{q}}^{\dag})$ and 
the complex bosons $(q,\widetilde{q}^{\dag})$
	\beq
	\begin{array}{ccc}
	& \psi_{q} & \\
	q & & \widetilde{q}^{\dag} \\
	& \psi_{\widetilde{q}}^{\dag} & 
	\end{array}
	,\eeq
but we will consider the theory dictated by only the chiral multiplet 
for the moment. 

For the description of a theory with supersymmetry, 
it is convenient to consider a field (super field) on 
super space which is the space with usual space-time real coordinates 
$x_{\mu}$ and the Grassman coordinates $\theta_{\alpha}$ ($\alpha =1,2$). 
In the present case of $N=2$ chiral multiplet, since it can be seen that 
it consists of two $N=1$ multiplets $(A_{\mu},\lambda )$ and 
$(\psi ,\phi )$ in view of $N=1$ supersymmetry, the Lagrangian of 
the $N=2$ supersymmetric \ym theory is compactly written as 
	\beq
	\L =\frac{1}{8\pi}\mbox{Im tr}\left[\tau_0 \int 
	d^2 \theta W^{\alpha}W_{\alpha} 
	+2\int d^2 \theta d^2 \overline{\theta}\Phi^{\dag}e^{-2V}
	\Phi\right]
	\lab{24}
	\eeq
by using the super fields 
	\beq
	W_{\alpha}=-i\lambda_{\alpha}-\frac{i}{2}
	(\sigma^{\mu}\bar{\sigma}^{\nu}
	\theta)_{\alpha}F_{\mu\nu} +\cdots ,\ 
	\Phi =\phi +\sqrt{2}\theta \psi +\cdots 
	,\eeq
where $\sigma^{\mu}=(\mbox{\boldmath$1$},\sigma^i )$, 
$\mbox{\boldmath$1$}$ is an unit matrix of the size $2\times 2$, 
$\sigma^i(i=1,2,3)$ are the Pauli matrices, 
$F_{\mu\nu}=\pa_{\mu}A_{\nu}-\pa_{\nu}A_{\mu}-i [A_{\mu},A_{\nu}]$. 
Ellipses are omission. $V$ is 
expressed by superfields, but the detail is not necessary here. 
In (\ref{24}), the complexified coupling constant 
	\beq
	\tau_0 =\frac{\theta_0}{2\pi}+i\frac{4\pi}{g^2}
	\lab{25}
	,\eeq
where $\theta_0 \in \mbox{\boldmath$R$}$ is the vacuum angle and 
$g$ is the gauge coupling constant, is introduced. 
The introduction of the vacuum angle is an analogy of 
$N=0$ (non-supersymmetric) QCD, where the vacuum angle is 
related to strong CP problem. Actually, $\theta_0$ in this 
$N=2$ theory can be set to zero because of chiral rotation of fermions. 
	
From the Grassman integral and expansion of the products of superfields, 
$\L$ becomes 
	\beq
	\L =\frac{1}{g^2}\mbox{tr\,}\left[-\frac{1}{4}F_{\mu\nu}F^{\mu\nu}
	+\frac{g^2\theta_0}
	{32\pi^2}F_{\mu\nu}\widetilde{F}^{\mu\nu}-\frac{1}{2}
	[\phi^{\dag},\phi]^2 +\cdots \right]
	\lab{26}
	,\eeq
where $\widetilde{F}^{\mu\nu}$ is the dual field strength 
defined by using the anti-symmetric tensor $\epsilon^{\mu\nu\rho\sigma}$ 
($\epsilon^{0123}=+1$) 
	\beq
	F^{\mu\nu}=\frac{1}{2}\epsilon^{\mu\nu\rho\sigma}
	F_{\rho\sigma}
	.\eeq

\begin{center}
\subsection{Vacuum }
\end{center}

In field theories, 
a complex scalar field often plays a role of Higgs field, and 
it can be shown that the present case is also the case. 
From (\ref{26}), it is easy to see that $\phi$ has the potential 
	\beq
	U=\frac{1}{2g^2}[\phi^{\dag},\phi ]^2 \ \underline{\underline{>}}0
	\lab{28}
	,\eeq
and the vacuum of the $N=2$ \ym theory is characterized by the minimum 
of this potential. In the case at hand, 
the vacuum corresponds to $U=0$, and $\phi =0$ can be considered 
as the candidate of $\phi$ which realizes the vacuum configuration. 
However, $U=0$ can be realized by some $\phi$ such that the 
commutator vanishes, so it is sufficient to parameterize it by 
using the generators $H_i$ of Cartan subalgebra of the 
gauge group $G$ \cite{KLYT,KLT,APS,HO} 
	\beq
	\phi =\sum_{i=1}^{\mbox{\scriptsize rank }(G)}a_i H_i
	,\eeq
where $a_i$ are complex parameters. 
Since the Weyl symmetry still remains in this parameterization, 
it would be convenient to characterize the theory by using 
quantities which are invariant under this symmetry. 
For example, in the case of SU($N_c$) gauge group the 
candidates are given by 
	\beq
	u_k =\mbox{tr }\langle \phi^k \rangle ,\ k=2,\cdots , N_c
	.\eeq
There exist more convenient parameters than $u_k$, defined by 
the symmetric polynomials 
	\beq
	s_k =(-1)^k \sum_{i_1 <\cdots <i_k =1}^{N_c}a_{i_1}\cdots 
	a_{i_k},\ k=2,\cdots,N_c
	,\eeq
but $\{ u_k\}$ and $\{ s_i\}$ are related each other by the 
Newton's formula
	\beq
	ks_k +\sum_{i=1}^{k}s_{k-i}u_i =0,\ s_0 =1, s_1 =u_1 =0
	.\eeq
$s_k$ are called moduli of the theory and the space of $s_k$ is called 
moduli space. 

\begin{center}
\section{Effective action and prepotential of SU(2) gauge theory}
\end{center}

\renewcommand{\theequation}{3.\arabic{equation}}\setcounter{equation}{0}

\begin{center}
\subsection{Effective action}
\end{center}

When $\phi$ has a non-zero vacuum expectation value, 
gauge fields gain masses by Higgs mechanism. For example, 
in the case of SU(2) gauge group, $\phi$ can be written by 
$\phi =a\sigma^3$, where $a =\langle \phi \rangle \in \mbox{\boldmath$C$}$, and 
the gauge symmetry breaks down from SU(2) to U(1) for $a\neq 0$. 
Then $W_{\mu}^{\pm}:=(A_{\mu}^1 \pm A_{\mu}^2 )/2$ and their 
supersymmetric particles (superpartners) gain mass $\sim a$, 
while $A_{\mu}^3$ and their superpartners remain massless. 
Furthermore, when the vacuum expectation value of 
$\phi$ is zero, the original SU(2) gauge symmetry 
is restored, therefore the classical moduli space 
has a singularity at $u\equiv\mbox{tr }\langle \phi^2 \rangle =2a^2 =0$.

On the other hand, in a quantum theory, the dynamics of particles is 
described by an action including quantum effects called effective action. 
In the case of $a\neq 0$, for large $a$ the masses of 
$W_{\mu}^{\pm}$ and corresponding superpartners also become large. 
These heavy particles do not ``actively'' move  and only massless particles 
play an essential role. Then these massive particles can be dropped out 
in the case that the energy scale, say $\l$, 
is small ($|a|\gg\l$). If these massive particles are integrated out 
in the path integral, only the light particles like $A_{\mu}^3$ will be left. 
The action obtained in this way is called low energy effective action 
(below, simply called effective action) and 
the theory is said to be in weak coupling region.

Actually, the explicit form of the effective action is expected to 
be complicated due to quantum fluctuation, but because of the $N=2$ 
supersymmetry this effective action(Lagrangian) is known to be simply represented by 
	\beq
	\L_{eff}=\frac{1}{4\pi}\mbox{Im}\left[\int d^2 \theta d^2 
	\overline{\theta}
	\frac{\pa \F(A)}{\pa A}\overline{A} 
	+\frac{1}{2}\int d^2 \theta d^2 \theta 
	\frac{\pa^2 \F(A)}{\pa A^2}W^{\alpha} W_{\alpha}\right]
	\lab{leff}
	,\eeq
where 
	\beq
	W_{\alpha}=-i\lambda^3 + \cdots ,\ 
	A=\phi^3 +\sqrt{2}\theta \psi^3 +\cdots
	\eeq
are the $U(1)$ multiplets and $\phi^3$ etc are the third component of 
$\phi$ etc. $\F (A)$ is a holomorphic function called prepotential 
satisfying $\pa \F (A)/\pa \overline{A}=0$, and it's second order derivative 
	\beq
	\tau =\frac{\pa^2 \F (A)}{\pa A^2}
	\lab{ttta}
	\eeq
is called as effective coupling constant. The lowest order term 
of (\ref{ttta}) is (\ref{25}). Furthermore, 
it is convenient to introduce 
	\beq
	A_{D} :=\frac{\pa\F}{\pa A}
	,\lab{ddi}
	\eeq
especially, for a later purpose. 

Note that the effective action is generated by the prepotential 
and once it is determined the quantum dynamics of the particles 
will be clarified. 
However, we are interested in the vacuum configuration of the theory, 
so $A$ can be replaced by it's scalar component $a$. 
Then $\F (A)$ reduces to $\F (a)$ and in this case (\ref{ddi}) 
is replaced by 
	\beq
	a_D : =\frac{\pa\F }{\pa a}
	\lab{dua}
	.\eeq
$\F (a)$ is a solution to the problem 
how to determine the low energy dynamics and 
the determination of $\F (a)$ is the subject of the discussions below.

\begin{center}
\subsection{Prepotential}
\end{center}

In the classical theory, the prepotential is given by 
	\beq
	\F_{cl}=\frac{\tau_0}{2}a^2
	\lab{cl}
	,\eeq
which is available from (\ref{25}) and (\ref{ttta}). 
On the other hand, perturbative part of the quantum prepotential 
can be determined from the beta function for coupling constant. 
In a gauge theory, coupling constant is not simply a constant, 
but is a quantity receiving quantum corrections and is 
determined by the beta function of renormalization group. 
In the case at hand, the beta function at 1-loop level is given by 
	\beq
	\mu\frac{dg}{d\mu}=\beta ,\ \beta =-bg^3 , \ b=\frac{1}{4\pi^2} 
	\lab{kuri}
	,\eeq
where $\mu$ is the renormalization scale, but since there is the 
non-renormalization theorem which sates that 
there are not corrections beyond 1-loop in perturbation 
theory \cite{HST,HSW}, (\ref{kuri}) exactly holds in this sense. 
Solving (\ref{kuri}) by imposing the condition $(g(a),\mu=a)$, one finds 
	\beq
	\frac{1}{g(a)^2}=\frac{1}{g^2}+2b\ln \frac{a}{\mu}
	,\eeq
where $g=g (\mu)$ at $\mu$. Therefore, 
	\beqa
	g(a)^2 &=&\frac{1}{\displaystyle \frac{1}{g(\mu)^2}+2b\ln
	\frac{a}{\mu}}\nm\\
	&\equiv& \frac{1}{\displaystyle 2b\ln\frac{a}{\l}}
	,\lab{38}
	\eeqa
where $\l$ is identified by 
	\beq
	\l\equiv\mu e^{-1/(2bg(\mu)^2)}
	.\eeq
$\l$ introduced in this way is called QCD (scale) parameter. 
Accordingly, the 1-loop prepotential can be determined from 
	\beq
	\frac{d^2\F}{da^2}\sim i\frac{4\pi^2}{g(a)^2}
	\eeq
as 
	\beq
	\F_{1-loop}=i\frac{a^2}{\pi}\ln \frac{a}{\l}
	.\lab{1-loop}
	\eeq

\begin{center}
\subsection{Instanton effect and prepotential}
\end{center}

Here, let us consider instantons and introduce the 
(Euclidean) action of \ym fields
	\beq
	S=-\frac{1}{4g^2}\int d^4 x \mbox{tr }(F_{\mu\nu}F^{\mu\nu})
	\lab{instac}
	.\eeq
Now, consider a configuration of gauge fields at 
infinity $(x\rightarrow \infty)$. Then $F_{\mu\nu}=0$ must 
be satisfied at infinity. This means that the gauge field tends to 
an equivalent configuration to the vacuum. 
The solution to classical 
equations of motion satisfying this condition is known as 
instanton solution. 
With the aid of this instanton solution (\ref{instac}) becomes 
	\beq
	S=\frac{8\pi^2}{g^2}
	,\eeq
and therefore the amplitude of instanton of 
unit topological charge is given by $e^{-S}$. 

From (\ref{38}) comparing the QCD parameter and the instanton amplitude, 
one finds that $(\l/a)^{4k}$ corresponds to the amplitude of 
instanton with topological charge $k$ ($k$-instanton)
	\beq
	\left(\frac{\l}{a}\right)^{4k} =e^{-\frac{8\pi^2}{g(a)^2}k}
	.\eeq
This factor is not proportional to powers in the coupling constant, 
therefore this can be considered as non-perturbative effect. 

Seiberg \cite{Sei} conjectured that for the actual prepotential 
these instantons contributed and it's form was predicted by 
	\beq
	\F=i\frac{a^2}{\pi}\left[\ln\frac{a}{\l}-\sum_{k=0}^{\infty}
	F_k \left(\frac{\l}{a}\right)^{4k}\right]
	.\lab{seipre}
	\eeq
Note that (\ref{seipre}) takes the form of a sum of (\ref{cl}), (\ref{1-loop}) and 
instantons. $F_1 \neq 0$ was pointed by Seiberg \cite{Sei}, but it was not known whether 
general $F_k$ were 0 or not. However, Seiberg and Witten \cite{SW1,SW2} 
showed that $F_k$ could be exactly determined by using data of 
a Riemann surface. 

\begin{center}
\subsection{Strong coupling region}
\end{center}

So far we have concentrated on the region $|a|\gg\l$, but next, 
let us consider the case of small $a$. Taking $a$ to be small, 
one finds that $a$ will arrive at the region $|a|\sim \l$. This corresponds to 
$u\sim \pm \l^2$ in terms of moduli. 
In this case, the analysis of the theory is very complicated, but according to the 
detailed study of Seiberg and Witten \cite{SW1,SW2} it turned out 
that at $u=+\l^2$ magnetic monopole (a particle carrying only unit 
magnetic charge) 
becomes massless and at $u=-\l^2$ dyon (a particle carrying both 
electric and magnetic charges) becomes massless. This indicates that 
quantum mechanically the moduli space has three singularities 
at $u=\pm \l^2$ and $\infty$. Note that the classical singularity 
disappears in the quantum theory.

In the description 
of the effective action presented in the previous (sub)sections, only 
the massless gauge fields and their superpartners were concerned, but 
in the present case 
monopoles and dyons must be taken into account. This indicates that 
the theory is in strongly coupled region and in this 
region it is not easy to write down the effective action, but by using a 
notion of duality the theory in this strong coupling 
region can be mapped to a weakly coupled dual theory. As a result, 
it becomes possible to gain understandings on the 
original theory by studying weakly coupled dual theory without 
directly treating the strongly coupled original theory. 
Seiberg and Witten \cite{SW1,SW2} showed that this was in fact possible. 

\begin{center}
\subsection{Duality}
\end{center}

According to Seiberg and Witten's result, 
the duality group $\Gamma (2)$ which is a subgroup of 
SL(2,$\mbox{\boldmath$Z$}$)
	\beq
	\Gamma (2) =\left\{\left.
	\left(\begin{array}{cc}
	a&b\\
	c&d
	\end{array}\right)\right | a\equiv d\equiv 1
	\mbox{ mod }2,\ b\equiv c\equiv 0 \mbox{ mod }2
	\right\}
	\eeq
is derived from the monodromy properties of $a$ and $a_D$ at each 
singularity and then it acts for the effective coupling constant which 
satisfies
	\beq
	\mbox{Im } \tau >0
	\eeq
as
	\beq
	\tau \longrightarrow \frac{a\tau +b}{c\tau +d}
	,\ \left(
	\begin{array}{cc}
	a& b \\
	c & d \end{array}
	\right)\in \Gamma (2)
	.\eeq
This fact implies that $\tau$ is a modulus of Riemann surface of genus 
one. From these consideration, they conjectured the existence of a 
torus which satisfied this condition and 
identified the complex $u$-plane with the quotient 
space $H/\Gamma (2)$ of the upper half plane $H$.

Of course, it can be observed that one of the generators of $\Gamma (2)$
	\beq
	S:\ \left(\begin{array}{cc}
	0&1\\
	-1&0
	\end{array}\right)
	\eeq
causes the exchange of $a$ and $a_D$ 
	\beq
	S:\ \tau=\frac{da_D}{da} \rightarrow -\frac{1}{\displaystyle 
	\frac{da_D}{da}} \equiv \tau_D 
	,\eeq
therefore the strong coupling region can be mapped to 
the weak coupling region of the theory having $\tau_D$ as 
the effective coupling constant. 

\begin{center}
\section{The \sw solution}
\end{center}

\renewcommand{\theequation}{4.\arabic{equation}}\setcounter{equation}{0}

\begin{center}
\subsection{The SU(2) case}
\end{center}

Searching whether a Riemann surface which satisfies this property does 
exist or not, Seiberg and Witten \cite{SW1} found that it was given by 
	\beq
	y^2 =(x^2 -\l^4 )(x-u)
	\eeq
on local complex coordinates $(x,y)\in \mbox{\boldmath$C$}^2$. 

Furthermore, if $\tau$ is identified with the modulus of a Riemann 
surface of genus one (torus), it should be represented by a ratio 
of periods of a holomorphic 1-form $dx/y$ along 
$\alpha$- and $\beta$-cycles on the torus. 
Regarding $a$ and $a_D$ as functions in $u$
	\beq
	\tau =\frac{da_D /du}{da/du}
	,\eeq
one may insure that this implies
	\beq
	\frac{da}{du}=\oint_{\alpha}\frac{dx}{y},\ 
	\frac{da_D}{du}=\oint_{\beta}\frac{dx}{y}
	.\eeq
Accordingly, in this view point, $a$ and $a_D$ 
can be interpreted as periods 
	\beq
	a =\oint_{\alpha}\lambda_{SW},\ a_D =\oint_{\beta}\lambda_{SW}
	,\lab{ssw}
	\eeq
of the meromorphic 1-form 
	\beq
	\lambda_{SW}=\int\frac{dx}{y}du =\frac{\sqrt{2}}{8\pi}
	\sqrt{\frac{x-u}{x^2 -\l^4}}dx
	\lab{dif}
	,\eeq
where in the second equality the normalization factor is 
introduced. The 1-form $\lambda_{SW}$ obtained in this way is usually 
referred to \sw differential or \sw 1-form, and the approach based on 
Riemann surface is often called \sw solution or \sw theory. 

\begin{center}
\subsection{Other gauge group cases}
\end{center}

In the case of classical Lie gauge groups except SU(2), 
it is known that the \sw curves are written by 
hyperelliptic curves. The SU(2) case is the only exception, and the 
\sw curve can have two representations, i.e., elliptic and 
hyperelliptic types. In the case not including 
quark hypermultiplets, these curves and associated 
\sw differentials are explicitly represented as follows. 

$A_n =$SU$(n+1)$ $(n>0)$: \cite{KLYT,KLT,APS,HO,AAG}   
	\beq
	y^2 =W_{A_n}^2 
	-\l_{A_n}^{2(n+1)}
	,\ \lambda_{A_n}=\frac{x\pa_x W_{A_n}}{y}dx
	\lab{su(N)}
	,\eeq
where 
	\beq
	W_{A_n}=x^{n+1}-\sum_{i=2}^{n+1}s_i x^{n+1-i}
	.\lab{an}
	\eeq

$B_n =$SO$(2n+1)$ $(n>1)$: \cite{AAG,AS,DS1,Han} 
	\beq
	y^2 =W_{B_n}^2 -\l_{B_n}^{4n-2}x^2 
	,\ \lambda_{B_n}=\frac{W_{B_n}-x\pa_x W_{B_n}}{y}dx
	,\eeq
where 
	\beq
	W_{B_n}=x^{2n}-\sum_{i=1}^{n}s_{2i}x^{2(n-i)}
	.\eeq
 
$C_n =$Sp$(2n)$ $(n>1)$: \cite{AAG,AS}
	\beq
	x^2 y^2 =W_{C_n}^2 -\l_{C_n}^{4(n+1)},\ 
	\lambda_{C_n}=-\frac{\pa_x W_{C_n}}{y}dx
	,\eeq 
where 
	\beq	
	W_{C_n}=x^2 \left[x^{2n}-\sum_{i=1}^{n}s_{2i}x^{2(n-i)}\right]
	+\l_{C_n}^{2(n+1)}
	.\eeq

$D_n =$SO$(2n)$ $(n>2)$: \cite{AAG,Han,BL}
	\beq
	y^2 =W_{D_n}^2 -\l_{D_n}^{4(n-1)}x^4 ,\ 
	\lambda_{D_n}=\frac{2W_{D_n}-\pa_x W_{D_n}}{y}dx 
	,\eeq
where 
	\beq
	W_{D_n}=x^{2n}-\sum_{i=1}^{n}s_{2i}x^{2(n-i)}
	.\eeq
Note that the normalization factors of \sw 1-forms are ignored 
in each case. The list used here is that given in \cite{IS}.

These hyperelliptic curves can be compactified to a 
Riemann surface with appropriate genus by adding an infinity 
(see figure 1), and taking canonical (symplectic) 1-cycles on this surface and 
denoting them as $\alpha_i $ and $\beta_i$, one sees that 
the \sw periods for a theory in gauge group $G$ can be written by 
	\beq
	a_i =\oint_{\alpha_i}\lambda_G ,\ 
	a_{D_i} =\oint_{\beta_i}\lambda_G  
	\lab{syuki}
	.\eeq
It is convenient to summarize these periods by 
	\beq
	\Pi =\left(\begin{array}{c}
	a_{D_i}\\
	a_i
	\end{array}\right)
	.\eeq
$\Pi$ is called period vector. 
	\begin{figure}[h]
        \begin{center}
        \epsfile{file=torr.eps,scale=0.6}
        \caption{1-cycles on $A_n$ type \sw curve identified with 
	genus $n$ Riemann surface}
        \end{center}
        \end{figure}

Also in the case including quarks, 
similar hyperelliptic curves and \sw differentials can be constructed, 
but since the number of quarks which can be added is restricted from 
asymptotic freedom and the form of the curve differs according to 
the number of quarks. 
For a later convenience, let us write down the 
\sw curve and \sw differential for the case of SU($N_c$) 
$(N_f <N_c =n+1)$ 
gauge group with $N_f$ quarks of mass $m_i$ ($i=1,\cdots,
N_f$) \cite{HO,AF}
	\beq
	y^2 =W_{A_n}^2 -\l_{A_n}^{2N_c -N_f}G,\ G=\prod_{i=1}^{N_f}(x+m_i )
	,\ \lambda=\frac{xdx}{y}
	\left(\frac{W_{A_n}\pa_x G}{2G}-\pa_x W_{A_n}\right)
	,\lab{con}
	\eeq
where $W_{A_n}$ is the simple singularity given in (\ref{an}). 
Note that $\lambda$ has poles at $x=-m_i$. 

{\bf Remark:} Also in the case of exceptional Lie gauge groups, 
\sw curves are known to be written in terms of 
hyperelliptic curves \cite{DS2,AAM,LPG}.
	
Below, the suffix of the QCD parameter is ignored, but this will 
not cause any confusion. 

\begin{center}
\section{\pf equations}
\end{center}

\renewcommand{\theequation}{5.\arabic{equation}}\setcounter{equation}{0}

\begin{center}
\subsection{SU(2) case }
\end{center}

As an example, let us consider the periods in the SU(2) gauge theory. 
These periods have been introduced by (\ref{ssw}), 
but in order to evaluate them it is necessary to explicitly express 
1-cycles as appropriate integral intervals. The simplest 
one to specify these intervals is to use 
branching points of \sw curve as 
	\beq
	\alpha :\ -\l^2 \longrightarrow +\l^2 ,\ 
	\beta :\ +\l^2 \longrightarrow u
	\eeq
and to recognize these as loops running counterclockwise (but $\alpha \cap \beta =+1$). 
Actually, since the 
evaluation of periods depends on the behavior of cycles, it is complicated 
to treat in general. For example, considering the calculation of periods at 
weak coupling 
region, one finds that the QCD parameter approaches to zero and $u$ 
moves to the other branching point, so the torus collapses as a result 
(see figure 2). 
Therefore the period $a$ ultimately reduced to an integral within infinitesimal 
interval near the origin, so the 
evaluation is easy, but since the dual period $a_D$ reduces to an integral 
from 0 to $\infty$, if under this situation $a_D$ is calculated, the 
integrand also diverges for $u\rightarrow\infty$, so that 
$a_D$ finally involves logarithmical divergence. 
Since we encounter such situation even if the gauge group is any group, 
the calculation of dual period (in weak coupling region) 
is not easy in general. 
	\begin{figure}[h]
        \begin{center}
        \epsfile{file=sing.eps,scale=0.6}
        \caption{Torus in the weak coupling region}
        \end{center}
        \end{figure}

Then, how should we do? One of the simplest way to 
evaluate the periods is to realize these periods as solutions to differential 
equation. In fact, it can be shown that the derivatives are 
	\beqa
	& &\frac{d \Pi}{du}=-\frac{\sqrt{2}}{16\pi}\oint_{\gamma}\frac{dx}{y}
	,\nm\\
	& &\frac{d^2 \Pi}{du^2}=\frac{\sqrt{2}}{32\pi}\frac{1}{(\l^4 -u^2 )}
	\oint_{\gamma}\sqrt{\frac{x-u}{x^2 -\l^4}}dx,
	\lab{ssw2}
	\eeqa
where the 1-cycles (or their linear combinations) are summarized as $\gamma$ \cite{KLT}. 
The second equation in (\ref{ssw2}) indicates the existence of the differential equation 
	\beq
	\frac{d^2 \Pi}{d u^2}+\frac{\Pi}{4(u^2 -\l^4)}=0
	\lab{su(2)picard}
	\eeq
with regular singularities and such equations are in general referred to \pf equation 
(for a history of \pf equation, see \cite{Gray}). 
Note that the periods are now constructed as solutions to 
differential equation. 

\begin{center}
\subsection{General case}
\end{center}

Even if the gauge group is any group, in order to derive \pf equations, it is 
sufficient to express parameter 
derivatives of period by another parameter derivatives. In fact, Alishahiha 
wrote the general form of \pf equations for all classical Lie gauge groups \cite{Ali1}. 
For example, in the case of SU($n+1$) \cite{KLT,Ali1}, they are 
	\beqa
	& &\left[ (n+1)\pa_{s_2}\pa_{s_n} -\sum_{i=2}^{n}(n+1-i)s_i 
	\pa_{s_{n+1}}\pa_{s_{i+1}}\right]\Pi =0,\nm\\
	& &\left[k\pa_{s_{n+2-k}}-(n+1)\pa_{s_2}\pa_{s_{n-k}}+\sum_{i=2}^{n}
	(n+1-i)s_i \pa_{s_i}\pa_{s_{n+2-k}}\right]\Pi=0,\ k=0,\cdots,n-2\nm\\
	& &\left[1+\sum_{i=2}^{n+1}i(i-2)s_i\pa_{s_i}+
	\sum_{i,j=2}^{n+1}ij s_i s_j\pa_{s_i}\pa_{s_j}-
	(n+1)^2 \l^{2(n+1)}\pa_{n+1}^2\right]\Pi=0,\nm\\
	& &\left[\pa_{s_i}\pa_{s_j}-\pa_{s_p}\pa_{s_q}\right]\Pi =0,\ 
	i+j=p+q
	.\eeqa

Of course, since the derivation of \pf equations is mechanical, 
some algorithms suitable for computer program can be made. 
One day Klemm {\em et al.} \cite{KTS} derived differential equations 
in Landau-Ginzburg model by using a method in singularity theory, 
but the algorithm of Isidro {\em et al.} \cite{IMNS1} corresponds to the 
generalization of this construction by replacing 
the holomorphic 1-form by \sw differential on \sw curve.

\begin{center}
\subsection{Hypergeometric differential equations}
\end{center}

\pf equations arising in this supersymmetric \ym theory are 
often identified with well-known hypergeometric 
differential systems. Gaussian hypergeometric function (in multiple 
variables) is the function which reduces to 
the single variable hypergeometric function $_ 2 F_1$ when one of the variables is 
non-zero and the remainings are set to zero, and Gaussian hypergeometric 
differential equations are the equations satisfied by such Gaussian 
hypergeometric functions \cite{SK}. 

\pf equation of SU(2) theory is given in (\ref{su(2)picard}), but for 
this equation performing the transformation of variable
	\beq
	z=u^2 /\l^4
	\lab{z}
	\eeq
one finds that (\ref{su(2)picard}) reduces to Gauss's hypergeometric 
equation \cite{KLT}
	\beq
	z(1-z)\frac{d^2 \Pi}{dz^2}+\left(\frac{1}{2}-\frac{z}{2}\right)
	\frac{d\Pi}{dz}-\frac{\Pi}{16}=0
	\lab{su}
	.\eeq
On the other hand, in the case of the SU(3) theory \cite{KLT}, 
\pf equations are represented by a system of simultaneous 
partial differential equations 
	\beqa
	& &[(27\l^6 -4u^3 -27v^2 )\pa_{u}^2 -12u^2 v\pa_u\pa_v -3uv\pa_v -u
	]\Pi=0,\nm\\
	& &[(27\l^6 -4u^3 -27v^2 )\pa_{v}^2 -36uv\pa_u\pa_v -9v\pa_v -3
	]\Pi=0	
	\lab{su(3)PF}
	,\eeqa
where $u\equiv s_2$ and $v\equiv s_3$ are SU(3) moduli, but this system 
is known to be equivalent to the two variable hypergeometric system of 
Appell's $F_4$ \cite{SK,S,E,Exton,Kim}
	\beqa
	& &
	\left[\tx \left(\tx -\frac{1}{3}\right)-x\left(\tx +\ty -\frac{1}{6}
	\right)\left(\tx +\ty -\frac{1}{6}\right)\right]\Pi=0,\nm\\
	& &
	\left[\ty \left(\ty -\frac{1}{2}\right)-y\left(\tx +\ty -\frac{1}{6}
	\right)\left(\tx +\ty -\frac{1}{6}\right)\right]\Pi=0
	\lab{appell}
	\eeqa
by the transformation of variables
	\beq
	x=4u^3 /(27\l^6 ),\ y=v^2 /\l^6
	\lab{xy}
	.\eeq
$\tx =x\pa /\pa x$ and $\ty =y\pa /\pa y$ are Euler partial 
derivatives. 

As for the other rank two gauge groups, there is 
$B_2 =C_2$, whose \pf equations 
can be recognized as Horn $H_5$ \cite{MSS}. 
Note that $_2 F_1 , F_4$ and $H_5$ are all Gaussian.

\begin{center}
\subsection{Solutions to the \pf equation}
\end{center}

Let us consider solutions to the SU(2) \pf equation 
(\ref{su(2)picard}) or (\ref{su}) at weak coupling region 
($u=\infty$) \cite{KLT}. According to Frobenius's method, it turns out 
that because of degeneracy of the solutions to indicial 
equation the solutions involve logarithm ($z=u^2 /\l^4$)
        \beq
        \rho_1 =z^{1/4}\sum_{i=0}^{\infty}\frac{a_i }{z^i},\ \rho_2 =
        \rho_1 \log \frac{1}{z}+z^{1/4}\sum_{i=1}^{\infty}\frac{b_i }{z^i}
        \lab{sol}
        ,\eeq
where the first several coefficients are given by 
        \beq
        a_0 =1,\ a_1 =-\frac{1}{16},\ a_2 =-\frac{15}{1024},\ 
	a_3 =-\frac{105}{16384},\ a_4 =-\frac{15015}{4194304},\ 
        a_5 =-\frac{153153}{67108864}
        \eeq
and 
        \beq
        b_1 =\frac{1}{8},\ b_2 =\frac{13}{1024},\ b_3 =\frac{163}{49152},\ 
        b_4 =\frac{31183}{25165824},\ b_5 =\frac{74791}{134217728}
        .\eeq

The next task is to relate these solutions to the periods, but is 
easily done by simply consider appropriate linear combination of 
them. However, the combination must be uniquely fixed by computing 
lower order expansion of the periods. 
This manipulation is equivalent to give an initial condition 
for the Picard-Fuchs equation. In this way, 
the result follows
        \beq
        a =\frac{\l}{\sqrt{2}}\rho_1 ,\ a_D =i\frac{\l}{\sqrt{2}\pi}
        (-4+6\log 2)\rho_1 -i\frac{\l}{\sqrt{2}\pi}\rho_2 
        \lab{212}
	.\eeq

\begin{center}
\section{Prepotentials}
\end{center}

\renewcommand{\theequation}{6.\arabic{equation}}\setcounter{equation}{0}
\begin{center}
\subsection{The SU(2) prepotential}
\end{center}

Let us derive the $SU(2)$ prepotential. The prepotential 
is a function in $a$ and is available from (\ref{dua}). However, 
since $a_D$ is obtained as a function in $u$, one must consider 
$a_D =a_D (a)$ by eliminating $u$ from this as a first step. 
This can be realized by inversely solving $a=a(u)$ in (\ref{212})
        \beq
        u=2a^2+\frac{\l^4}{16a^2}+\frac{5\l^8}{4096a^6}+
        \frac{9\l^{12}}{131072a^{10}}+\cdots
        .\lab{gyaku}
	\eeq

Next, after substituting this expression into (\ref{dua}), expanding it 
for large $a$ and further integrating it over $a$, one finds the 
prepotential \cite{KLT}
	\beq
        \F =i\frac{a^2}{\pi}\left[
        \log \left(\frac{a}{\l}\right)^2 +
        4\log 2-3-\sum_{k=1}^{\infty}F_k \left(\frac{\l}{a}
        \right)^{4k}\right]
        ,\eeq
where $O(a)$-terms including integration constant are 
ignored because they do not affect to the effective coupling constant. 
The first two coefficients of the instanton expansion are given by 
        \beq
        F_1 =\frac{1}{64},\ F_2 =\frac{5}{32768}
        \lab{inexp}
        .\eeq

Since $F_1$ is $F_1 \sim 0.016$, it is a number near 0 and also 
$F_2$ is a very small number $F_2 \sim 0.00015$. From these values, 
the contributions from instantons are small quantities which 
may be ignored as a matter of fact. Nevertheless, they contributes 
non-zero effects. The calculation of these very small 
but non-zero non-perturbative effects by using the methods in 
field theory requires much labour, but the point that gave 
a systematic method which can calculate such effect was 
one of the brilliant success of Seiberg and Witten. 

\begin{center}
\subsection{The case including quarks}
\end{center}

So far we have considered pure \ym theory, that is, the theory 
not including quarks. However, since QCD not including quarks 
is physically unnatural, let us consider the prepotential 
with quarks. 

Again restricting the gauge group to SU(2), one can see that 
quarks can be added up to three, provided the asymptotic 
freedom is preserved. Seiberg and Witten \cite{SW2} conjectured that the 
prepotential including $N_f$ massless quarks was given by 
	\beq
	\F_{N_f}=\frac{ia^2}{\pi}\left[\frac{4-N_f}{4}\ln \left(\frac{a}{\l}
	\right)^2 +\sum_{i=0}^{\infty}F_k \left(\frac{\l^2}{a^2}\right)^{
	(4-N_f )i}\right]
	\lab{Nfpri}
	,\eeq
and the validity of this formula was later proved by 
Ito and Yang \cite{IY1} (for the analysis of periods, see \cite{Ryang}). 

Actually, since the masses of the quarks are expected to be non-zero, 
the prepotential in the theory including quarks with 
masses $m_i$ is known to be modified to \cite{Ohta} 
	\beqa
	\F_{N_f}&=&i\frac{\wa^2}{\pi}\left[\frac{4-N_f}{4}
	\ln\left(\frac{\wa}{\l}\right)^2 +\F_{0}^{N_f}-
	\frac{\sqrt{2}\pi}{4i\wa}\sum_{i=1}^{N_f}n_{i}^{\,'}m_i 
	+\frac{N_f}{2}\wa\right.\nm\\
	& &+\left.\frac{1}{4\wa^2}\left(\frac{3}{2}\sum_{i=1}^{N_f}m_{i}^2 -
	\F_{s}^{N_f}\right)+\sum_{i=2}^{\infty}
	F_i (\l^{4-N_f} ,m_1 ,\cdots ,m_{N_f} )\wa^{-2i}\right]
	\lab{massivepri}
	,\eeqa
where $n,n'\in \mbox{\boldmath$Z$}$ are winding numbers of 
1-cycles which enclose the poles of the \sw differential 
(c.f (\ref{con})) and 
	\beq
	\F_{0}^{N_f}=
	\sum_{i=1}^{N_f}\left(\wa -\frac{m_i}{\sqrt{2}}\right)^2 
	\ln \left(\wa -\frac{m_i}{\sqrt{2}}\right)+
	\sum_{i=1}^{N_f}\left(\wa +\frac{m_i}{\sqrt{2}}\right)^2 
	\ln \left(\wa +\frac{m_i}{\sqrt{2}}\right)
	.\eeq
Furthermore, $\F_{s}^{N_f}$ is some constant (corresponding to 
classical effective coupling constant) and $\wa$ is a quantity 
that the residue contribution of \sw 1-form is subtracted from $a$. 
The first two expansion coefficients of $N_f =1$ theory are given by 
	\beq
	F_2 =-\frac{\l^3 m_1}{64},\ F_3 =\frac{3\l^6}{16384}
	.\eeq
It is now easy to see the dependence of the masses of the quarks. 
(\ref{massivepri}) is complicated than (\ref{Nfpri}), but note that in 
(\ref{massivepri}) for $m_i \rightarrow 0$ it reduces to 
(\ref{Nfpri}), so (\ref{massivepri}) includes 
the Seiberg and Witten's formula (\ref{Nfpri}). 

\begin{center}
\subsection{Check}
\end{center}

The prepotential has been obtained in this way, but does it really 
have field theoretic meaning? The calculation presented so far is 
based on the data of a Riemann surface, but if it has a 
physical meaning, the validity of it must be discussed by a method of 
field theory. One can see that the prepotential up to one-loop level 
certainly coincided with the result of perturbative calculus, 
but the instanton effects, specifically, its expansion coefficients 
coincide with the value expected from field theory? 

As a method to check this, there is a complicated method 
called instanton calculus and the instanton contribution can be 
determined by this. 
However, since the actual calculation is very complicated, explicit 
calculation is usually 
proceeded up to 2-instanton level, but at least up to this level it is 
confirmed that the result based on Riemann surface is not contradict 
to the instanton calculus. 
In this way, Seiberg and Witten's approach gained supports that 
it is correct also as a physics \cite{IS,FP,IS3,Slat,DKM1,DKM3,AHSW,HS}. 

Though once it was pointed out that there was a contradiction with the 
result of instanton calculus in the case of SU(2) with three 
quarks, 
this was resolved by admitting a linear transformation which shifts moduli by 
QCD parameter \cite{DKM3,AHSW,HS}. 
The linear transformation mentioned here is to specify 
where is the origin of the moduli space, so this is trivial in a sense, 
but note that in order to derive instanton contribution to prepotential 
from \sw curve this shift is important. This kind of discrepancy seems to 
admit generally for the hyperelliptic curves including quarks constructed 
so far, in fact, also in the case of SU(3) with four and five quarks, 
it is observed by Ewen and F\"{o}rger that such constant shift of moduli is 
necessary to correctly include instanton effects \cite{EF2}. 

Moreover, when the gauge group is exceptional Lie groups, it is 
known that there is a contradiction between the results by hyperelliptic curves 
and by instanton calculus. In these cases, it is known that if particular 
spectral curves, non-hyperelliptic curves often referred to square root 
type \cite{MW,LW,EY2}, are regarded as \sw curves then the instanton contribution to 
the prepotentials from these curves coincide with the 
instanton calculus \cite{Ito,Ghe}. 
In this sense \sw curves should be formulated by spectral curves of 
integrable system rather than hyperelliptic curves not only in classical 
but also in exceptional gauge groups.

\begin{center}
\section{Analytic continuation of Period integrals}
\end{center}

We have overviewed the theory of prepotential from the view point of 
\pf equation. Next, let us consider prepotentials from other standpoint. 

\renewcommand{\theequation}{7.\arabic{equation}}\setcounter{equation}{0}
\begin{center}
\subsection{Period integrals}
\end{center}

We have introduced \pf equations in order to remove the labour of 
direct evaluation of periods because it is very intractable in general. 
As a matter of fact, it is 
sufficient to obtain solutions at regular singularities of \pf equations 
in order to study the behavior of periods at each singularities on 
the moduli space, but in the method intermediated by \pf equations, 
since the analysis depends on case by case such as SU(2) or SU(3), 
the derivation of general form of instanton correction terms is 
not easy. However, for about 1-instanton level, it can be slightly easily 
derived by using application of analytic 
continuation of period integrals \cite{MSS,MS1,DKP1,DP}, although 
the derivation of higher order instanton corrections in this 
method will not work well because of technical problems. 
This approach was taken in the analysis of periods of Calabi-Yau 
manifolds, and the case at hand can be regarded as a version of them. 

Let us consider the periods of SU($N_c$) gauge theory as an example, 
but details are omitted here. Since the \sw curve is $2N_c$-order polynomial, it has $2N_c$ 
zeros $e_i$ (branching points) on $x$-plane. 
In the weak coupling region, expanding the \sw differential around 
$\l\sim 0$ and then integrating it over $\alpha_i$-cycle which 
enclose $e_i$, then one gets 
	\beq
	a_i =e_i +\sum_{n=1}^{\infty}\frac{(1/2)_n \l^{2N_c n}}{n!(2n)!}
	\left(\frac{\pa}{\pa e_i}\right)^{2n-1}\prod_{k,k\neq i}
	(e_k -e_i )^{-2n}
	.\eeq
The calculation over $\beta_i$-cycle defined in a similar way 
involves logarithmical 
divergence. Then rewriting \sw differential as 
	\beq
	\lambda =dx\int_{-i\infty}^{i\infty}\frac{ds}{2\pi i}
	\frac{\Gamma (-s)\Gamma (s+1/2)}{2s\Gamma (1/2)}
	\prod_{k=1}^{N_c}(x-e_k )^{-2s}(-\l^{2N_c})^s
	\lab{ry}
	\eeq
and performing residue calculus over $s=\{ 0\} \cup \mbox{\boldmath$N$}$ and integrating it over $\beta_i$-cycle, 
one finds that $a_{D_i}$ is given by 
	\beqa
	a_{D_i}&=&\frac{i}{2\pi}\sum_k (a_i -a_k )\ln 
	\left(\frac{e_i -e_k}{\l}\right)^2 -\frac{i}{\pi}\sum_k 
	(e_i -e_k )+\frac{i\ln 2}{\pi N_c}\sum_k (a_i -a_k )\nm\\
	& &-i\frac{\l^{2N_c}}{8}\frac{\pa}{\pa e_i}\sum_k 
	\frac{1}{\prod_{l\neq k}(e_k -e_l )^2}
	.\eeqa
Rewriting this by $a_j$ and integrating it over the period, one can find 
the 1-instanton contribution to the prepotential and the result 
coincides with the instanton calculus. 

This approach to get a general form of prepotential, in particular, 
the instanton correction part, from period integrals was proceeded in 
\cite{MSS,MS1,DKP1,DP}, but 
the most characteristic advantage of this method was that the general 
form of prepotential could be derived both in weak and strong 
coupling region without using \pf equations. 
D'Hoker {\em et al.} \cite{DKP1} determined the prepotentials based on 
all classical Lie gauge groups and the result showed the agreement 
with the instanton calculus. Furthermore, D'Hoker and Phong \cite{DP} 
succeeded to give a formula of the prepotential 
in the strong coupling region for the SU($N_c$) gauge theories. 

\begin{center}
\section{The scaling relation}
\end{center}

\renewcommand{\theequation}{8.\arabic{equation}}\setcounter{equation}{0}

It is well-known that the prepotential satisfies a very 
helpful homogeneous relation (Euler equation) called scaling 
relation \cite{Mat}. In this section, the basics of the 
scaling relation are discussed. 

\begin{center}
\subsection{Transformation rule of prepotential}
\end{center}

Firstly, let us see how the prepotential transforms under the action of 
$\Gamma (2)$ according to Matone \cite{Mat}. We have already seen that 
the effective coupling constant transforms under the 
action of $\Gamma (2)$. Since the effective coupling constant 
is a second order derivative of the prepotential, this is equivalent to 
	\beq
	\displaystyle
	\frac{\pa^2 \wF (\wa )}{\pa \wa^2}=\frac{\displaystyle 
	A\frac{\pa^2 \F}{\pa a^2}
	+B}{\displaystyle C\frac{\pa^2 \F}{\pa a^2}+D}
	\lab{henk}
	,\eeq
where $\wa =C a_D +D$. The left hand side of (\ref{henk}) 
is written by $\wa$, but by rewriting this by $a$, one finds 
	\beq
	\frac{\pa^2 \wF (\wa )}{\pa \wa^2}=\left[-\left(\frac{\pa a}{\pa \wa}
	\right)^3 \frac{\pa^2 \wa}{\pa a^2}\frac{\pa }{\pa a}+
	\left(\frac{\pa a}{\pa \wa}\right)^2\frac{\pa^2 }{\pa a^2}
	\right]\wF (\wa )
	,\eeq
so from (\ref{henk}) 
	\beq
	(C \F^{\,''}+D)\pa_{a}^2 \wF -C\F^{\,'''}\pa_a \wF -
	(A\F^{\,''}+B)(C\F^{\,''}+D)^2 =0
	\lab{b}
	\eeq
can be obtained. Here, 
$\F^{\,''}=\pa_{a}^2 \F (a)$ and $\wF =\wF (\wa )$. 
(\ref{b}) can be solved over $\wF$ and in fact it has a solution 
	\beq
	\wF (\wa )=\F (a)+\frac{1}{2}(AC a_{D}^2 +BD a^2 )
	+BCa a_D +c (Ca_D +Da)
	,\eeq
where $c$ is the integration constant, but can be set to zero because 
the linear term concerning periods does not contribute to the 
effective coupling constant. In this way, the transformation rule 
is obtained. 

Next, considering the quantity 
	\beq
	\G (a) =\F -\frac{a}{2}\frac{\pa \F}{\pa a}
	, \eeq
one finds that $\G$ is invariant under the action of $\Gamma (2)$. 
Accordingly, $\G$ is a modular invariant. $\G$ is a function in 
$a$, but by regarding $a=a(u)$ ($'=d/du$), it follows that 
	\beq
	\frac{d\G}{du}=\frac{1}{2}(a^{\,'}a_D -aa_{D}^{\,'})
	\lab{m}
	.\eeq	

Now, look at the right hand side of (\ref{m}). Since 
it corresponds to the Wronskian of the \pf equation, it is actually a constant. 
To determine this constant, it is enough to substitute the periods obtained 
by \pf equation (\ref{212}) into this expression. Thus, 
	\beq
	\frac{d\G}{du}=-\frac{i}{2\pi}
	\lab{n}
	.\eeq
Therefore,  
	\beq
	\F -\frac{a}{2}a_D =-\frac{i}{2\pi}u
	\lab{78}
	.\eeq
This is the scaling relation of prepotential.

{\bf Remark:} After the discovery of this relation by Matone, 
it was checked that (\ref{78}) holds exactly in view of instanton 
calculus by Fucito and Travaglini \cite{FT}.

\begin{center}
\subsection{Other view points}
\end{center}

After the discovery of Matone the same scaling relation 
was rederived in various view points. For example, there are 
the degree counting of the effective coupling constant \cite{STY}, 
Whitham theory of soliton \cite{EY}, 
anomalous superconformal Ward identity \cite{HW} 
and direct derivation by \pf equation \cite{KO}.

Here let us derive the scaling relation for SU($N_c$) as an example. 
Firstly, recall the \sw curve given in (\ref{su(N)}). For the 
variables in this curve, associating degree (mass dimension)
	\beq
	\mbox{deg }(y):\mbox{deg }(x):\mbox{deg }(s_i ):\mbox{deg }(\l)
	=n+1 :1:i:1
	\eeq
one can see that this curve is homogeneous. On the other hand, since the 
periods are determined from \sw differential, that the 
periods found to have degree 1 in a similar way.

Accordingly, periods can be seen as a homogeneous function in 
moduli and QCD scale parameter, and as a result the Euler equations 
are derived to satisfy 
	\beq
	a_i =\sum_{j=2}^{N_c}s_j\frac{\pa a_i}{\pa s_j}+
	\l \frac{\pa a_i}{\pa \l},\ a_{D_i}=
	\sum_{j=2}^{N_c}s_j\frac{\pa a_{D_i}}{\pa s_j}+
	\l \frac{\pa a_{D_i}}{\pa \l}
	\lab{eu}
	.\eeq
Since $a_{D_i}$ must be expressed by $a_j$ when 
we think of prepotential, and by interpreting $a_{D_i}=a_{D_i}(a_j (s_k ,\l ),\l)$ 
and using the second equation in (\ref{eu}) to the first equation, one gets 
	\beq
	a_{D_i}=
	\sum_{j=1}^{N_c}a_j\frac{\pa a_{D_i}}{\pa a_j}+
	\l \frac{\pa a_{D_i}}{\pa \l}
	\lab{eu2}
	,\eeq
where $\pa /\pa \l$ acts only the second argument of $a_{D_i} =(a_j ,\l)$. 
Substituting (\ref{dua}) into (\ref{eu2}) and further integrating it 
over $a_i$, one gets 
	\beq
	\sum_{i=1}^{N_c}a_i \frac{\pa \F}{\pa a_i}+\l\frac{\pa \F}{\pa \l}
	-2\F =0
	\lab{scaling}
	,\eeq
where integration constants are ignored because it can be 
absorbed by redefinition of prepotential. (\ref{scaling}) is the one 
known as scaling relation of prepotential.

Actually, it is known that $\l \pa \F/\pa \l$ is proportional to the 
product of the coefficient of 1-loop beta function and the 
moduli \cite{EY}, therefore in the case of SU(2) the scaling relation 
is given by 
	\beq
	aa_D -2\F =\frac{i}{\pi}u
	\lab{ko}
	,\eeq
which coincides with (\ref{78}).

\begin{center}
\subsection{Scaling relation including massive quarks}
\end{center}

Next, let us consider the scaling relation 
of prepotential of the case including quarks. 
According to the result, in the case of SU($N_c$) with $N_f$ quarks of 
mass $m_i$, the prepotential satisfies 
	\beq
	2\F-\sum_{i=2}^{N_c}a_i \frac{\pa \F}{\pa a_i}=
	\sum_{i=2}^{N_f}m_i \frac{\pa \F}{\pa m_i}+\l\frac{\pa \F}{\pa \l}
	\lab{sc}
	,\eeq
but in the case that the masses are not all zero (\ref{sc}) does not 
give a simple relation between moduli, quark masses and 
prepotential in contrast with the previous example. 
However, the right hand side of (\ref{sc}) can be calculated by 
another method, and in the case of SU(2) with $N_f =1$ massive 
quark of mass $m$ the scaling relation takes the form 
($' =d/du$) \cite{Ohta3}
	\beqa
	a\frac{\pa \F}{\pa a}-2\F &=&\frac{m}{2\sqrt{2}}(n^{\,'}a-na_D )-
	\frac{m^2}{16}\left(i+4i\ln 2 -2\pi nn'\right)\nm\\
	& &-\frac{1}{4}\int \left[a_{D}^{\,'}\int 
	\frac{a^{\,'}Z}{4m^2 -3u}du-a^{\,'}\int \frac{a_{D}^{\,'}Z}{
	4m^2 -3u}du\right]du 
	\lab{sc2}
	,\eeqa
where 
	\beq
	Z=-8m^2 +3\frac{3m\l^3 -4u^2}{4m^2 -3u}
	\eeq
and $n,n^{\,'}\in \mbox{\boldmath$Z$}$ are the winding numbers of 
1-cycles around the pole $x=-m$ of \sw differential.

\begin{center}
\subsection{Examples of prepotentials obtained by scaling relation}
\end{center}

In general, the calculation of prepotential in higher rank gauge group 
(with massive quarks) is very complicated, but instanton 
coefficients can be directly determined if the scaling relation is used. 
The SU(2) example is presented in appendix B. 
As for the other examples of prepotentials calculated in this approach, 
there are: 
SU(3) with quarks \cite{EF2,EF1}, $G_2$\cite{Ito} and $E_6$ \cite{Ghe}.

\begin{center}
\section{WDVV equations and perspective}
\end{center}

\renewcommand{\theequation}{9.\arabic{equation}}\setcounter{equation}{0}
\begin{center}
\subsection{WDVV equations}
\end{center}

We have seen that the prepotential satisfies a relation 
called scaling relation, but it also satisfies another 
important relation. It is the Witten-Dijkgraaf-Verlinde-Verlinde 
(WDVV) equations, often appears in two-dimensional topological field 
theory \cite{Dub}. 

In two-dimensional topological field theory, the topological free 
energy $F=F(t^1 ,\cdots ,t^n )$, where $t^i$ are flat coordinates, 
which is a generating function of 
correlation functions, has following properties. 

For the third order derivative
	\beq
	c_{ijk}(t):=\frac{\pa^3 F}{\pa t^i \pa t^j \pa t^k}
	,\eeq
it satisfies 
	\beq
	\eta_{ij}:=c_{1ij}(t)=\frac{\pa^3 F}{\pa t_1\pa t_i \pa t_j}=
	\mbox{constant},\ 
	\eta^{ij}:=(\eta_{ij})^{-1},\ c_{ij}^k :=\eta^{kl}c_{lij}(t)
	\eeq
and from commutativity of the structure constant $c_{ijk}(t)$, the 
WDVV equations follow
	\beq
	F_{ijk}\eta^{kl}F_{lmn}=
	F_{njk}\eta^{kl}F_{lmi},\ F_{ijk}\equiv \frac{\pa^3 F(t)}{\pa t^i 
	\pa t^j \pa t^k}
	.\lab{93}
	\eeq

On the other hand, the "WDVV equations" in four-dimensional $N=2$ \ym theory of gauge group $G$ look like (\ref{93}) but take 
the form \cite{IY3,MMM1,MMM3,IXY}
	\beq
	(\F_i )(\F_k )^{-1}(\F_j )=(\F_j )(\F_k )^{-1}(\F_i ),\ 
	i,j,k =1,\cdots,\mbox{rank}(G)
	\lab{wdvv2}
	,\eeq
where the symbols mean the matrix notation 
	\beq
	(\F_i )_{jk} =\frac{\pa^3 \F}{\pa a_i \pa a_j \pa a_k}
	\lab{81}
	.\eeq

Marshakov {\em et al.} \cite{MMM1,MMM3} pointed out that 
the WDVV equations in this form are satisfied by not only the 
prepotentials in four-dimensional $N=2$ \ym theories 
but also those in five-dimensional supersymmetric \ym theories 
of $S^1$ compactification models. 
This indicates that these two kinds of theories are 
simply one of solutions to the WDVV equations (\ref{wdvv2}). Accordingly, 
the manipulation to get prepotential extensively stated in the 
previous sections was simply a labour to obtain one of the solutions 
to (\ref{wdvv2})! For this reason, in order to gain understandings on 
these supersymmetric \ym theories, it is necessary to study them 
in the frame work of WDVV equations.

\begin{center}
\subsection{Perspective}
\end{center}

In view of WDVV equations, the low energy effective theory of 
four-dimensional supersymmetric \ym theory seems to imply that 
it has a nature as a topological field theory, but since this effective 
theory is not necessary to be topological field theory, 
it is interesting that the equations (\ref{wdvv2}) like (\ref{93}) hold. 
However, it is not still turned out whether this effective theory 
can be really interpreted as topological field theory. 
Moreover, the WDVV equations in this form is known to widely hold 
also in the case including massive quarks, and in such case the masses of 
quarks are regarded as if they are one of periods \cite{MMM3}. 
In this sense, the origin of the mass of the quarks may be explained 
in the frame of topological field theory. 

Nevertheless, since the meaning and importance of the role of 
(\ref{wdvv2}) in this supersymmetric \ym theory are not still 
clarified, studying (\ref{wdvv2}) in detail may provide something new 
aspect of $N=2$ supersymmetric \ym theory. 

\begin{center}
\section*{Acknowledgment}
\end{center}

I acknowledge prof. Takahiro Kawai for giving me the 
opportunity as a talker of the colloquium.

\begin{center}
\section*{A. Other representation of \pf equations }
\end{center}
\renewcommand{\theequation}{A.\arabic{equation}}\setcounter{equation}{0}

Usually in most cases, \pf equations are represented by using 
moduli derivatives, but in the cases except SU(2) are 
represented by a system of simultaneous 
partial differential equations. 
However, these \pf equations can be expressed into a more 
convenient form. 

According to the result \cite{Ohta2}, 
it is better to make differential equation by 
using QCD scale parameter. This is because \sw differentials in any 
(classical) gauge group always take the form
	\beq
	\frac{d^k \lambda}{d \l^k}=\frac{\mbox{polynomial in }x}{y}dx
	\eeq
and this quantity always representable by a linear combination of 
Abelian differentials \cite{KTS}, and therefore summing up over various $k$ produces 
ordinary differential equation \cite{Ohta2}. However, the equation must have periods as 
independent solutions, so the order of the equation is the same with the total 
number of periods. 

Let us cite the SU(3) \pf ordinary differential equation \cite{Ohta2} 
for a reference. Denoting the equation by ($z=\l^6$)
	\beq
	\frac{d^4 \Pi}{d z^4}+\sum_{i=0}^{3}c_i \frac{d^i \Pi}{d z^i}=0
	\lab{ode}
	,\eeq
one finds that the coefficients $c_i$ are 
	\beqa
	c_0&=&{\frac{-45 \left( 3 z - 4 {u^3} + 27 {v^2} \right)}
  	{2 {z^2} \widetilde{\Delta}_{SU(3)} }},\nm\\
	c_1&=&{\frac{45 \left( 1053 {z^2} - 538 z {u^3} + 40 {u^6} + 
    	3267 z {v^2} - 54 {u^3} {v^2} - 1458 {v^4} \right) }
	{2 {z^2} \widetilde{\Delta}_{SU(3)}}},\nm\\
	c_2&=&\frac{1}{4 {z^2} \widetilde{\Delta}_{SU(3)}}\left[
	445905 {z^3} - 8 \left(4 {u^3}-
	 27 {v^2}\right)^3+{z^2} \left(-217368 {u^3}+734589 {v^2}
	\right)\right.\nm\\
	& &\left.+ 36 z \left( 676 {u^6} - 135 {u^3} {v^2} - 
        29889 {v^4} \right) \right],\nm\\
	c_3&=&\frac{1}{z \widetilde{\Delta}_{SU(3)}}
	\left[76545 {z^3} - 162 {z^2} 
      	\left(244 {u^3}-297{v^2} \right)  - 
     	4 {{\left( 4{u^3}-27{v^2} \right) }^3}\right.\nm\\
	& &\left. +9z\left( 656 {u^6}-1080{u^3}{v^2}- 
        22599{v^4} \right)\right]
	,\eeqa
where $\widetilde{\Delta}_{SU(3)}$ is the product 
	\beq
	\widetilde{\Delta}_{SU(3)}=(15 z - 4 u^3 + 27 v^2)
	\Delta_{SU(3)}
	\eeq
of the discriminant of SU(3) \sw curve 
	\beq
	\Delta_{SU(3)}=\left[729 z^2 + \left(4 u^3 - 27 v^2\right)^2 - 
        54 z \left(4 u^3 + 27 v^2\right)\right]
	.\eeq

At first sight, this ordinary differential system seems to be more complicated 
than the original SU(3) \pf system (\ref{appell}), 
but it has an advantage that the solutions take a convenient form. 

The moduli space of the SU(2) theory 
can be interpreted as a Riemann sphere with three singularities (after suitable 
compactification), but that of the SU(3) theory can be seen as 
the complex projective space $\mbox{\boldmath$CP$}^2$ \cite{KLT}, which 
can be covered by the three coordinate neighborhood 
	\beq
	P_1 :\ \left(\frac{4u^3}{(27\l^6 )^2}:\frac{v^2}{27\l^{12}}:1\right)
	,\ P_2 :\ \left(\frac{4u^3}{v^2}:1:\frac{27\l^{12}}{v^2}\right),\ 
	P_3 :\ \left(1:\frac{27v^2}{4u^3}:\frac{(27\l^6 )^2}{4u^3}\right)
	.\eeq
We can choose $P_2$ and $P_3$ as weak coupling region. We faced on a similar situation when we discussed the mirror symmetry 
of Calabi-Yau manifold with several complex structure moduli. There, how to choose the large radius limit was a problem, 
so some people might remember this. In the case at hand, there are two choices, but 
in view of QCD parameter both $P_2$ and $P_3$ can be thought to be in the 
region $\l\sim 0$. Thus the solutions to (\ref{ode}) give common 
basis on these two coordinates.

\begin{center}
\section*{B. Application of scaling relation}
\end{center}
\renewcommand{\theequation}{B.\arabic{equation}}\setcounter{equation}{0}

Since the scaling relation is simply an Euler equation, 
it is not so interesting at first sight, but let us 
show that this relation is very useful by taking 
a derivation of the SU(2) prepotential as an example \cite{Mat}.

Firstly, let us recall that the prepotential was determined 
by calculating the periods by \pf equation. 
In that calculation, the period $a$ was a function in the moduli $u$, 
but $u$ was finally represented by $a$. What happens, if 
this calculation is applied to \pf equation (\ref{su(2)picard})?
(\ref{su(2)picard}) is written by $u$ derivative of $a$, 
but regarding $u$ as a function in $a$, i.e., $u=u(a)$, 
one finds ($' =d/da$)
	\beq
	\frac{da}{du}=\left(\frac{du}{da}\right)^{-1}=u^{\,'-1},\ 
	\frac{d^2 a}{du^2}=-u^{\,'-3}u^{\,''}
	\lab{bibun}
	.\eeq
Substituting this into (\ref{su(2)picard}), one can arrive at the 
cerebrated Matone's differential equation \cite{Mat}
	\beq
	4(\l^4 -u^2 )u^{\,''}+au^{\,'3} =0
	\lab{mato}
	.\eeq
This equation enables to determine $u$ as a function in $a$, if 
$u$ is solved over $a$ (\ref{gyaku})
	\beq
	u=\sum_{i=0}^{\infty}f_i a^{2-4i}
	\lab{717}
	,\eeq
where the expansion coefficients are denoted by $f_i$ with $f_0 =2$.

Here, let us recall the scaling relation (\ref{78}). Since up to 1-loop 
level prepotential can be determined from perturbation theory, this 
part may be thought as already known, although 
the general form of prepotential is always written in the form 
(classical)$+$(1-loop)$ + $(instantons). For this, regarding only the instanton expansion 
coefficients are unknowns and setting 
	\beq
	\F=i\frac{a^2}{\pi}\left[\ln \left(\frac{a}{\l}\right)^2
	-\sum_{i=0}^{\infty}\F_i \left(\frac{\l}{a}\right)^{4i}
	\right]
	\eeq
and substituting this and (\ref{717}) into (\ref{78}), one finds 
	\beq
	f_k =4k\F_k 
	.\eeq
As was already stated, the calculation of $a_D$ is very 
complicated, but this method directly determine the expansion coefficients. 

Moreover, since (\ref{78}) directly relates the moduli and prepotential, 
the differential equation satisfied by prepotential  
	\beq
	\F^{\,'''}=\frac{\pi^2}{4}\frac{(a\F^{\,''}-\F^{\,'})^3}
	{\l^4 +\pi^2 (a\F^{\,'}-2\F)^2}
	\eeq
is available, provided $u$ of (\ref{mato}) is substituted into 
$u$ of (\ref{78}). 

{\bf Remark:} The method of scaling relation played an important role 
also in the proof \cite{KO} of 
Nekrasov's insist that states the \sw theory can be obtained from 
a circle compactification of five-dimensional supersymmetric \ym theory \cite{Nek}.

\begin{center}

\end{center}

\begin{thebibliography}{99}

\bibitem{SW1}
N. Seiberg and E. Witten, 
Nucl. Phys. B {\bf 431}, 484 (1994). 

\bibitem{SW2}
N. Seiberg and E. Witten, 
Nucl. Phys. B {\bf 435}, 129 (1994).

\bibitem{WB}
J. Wess and J. Bagger, 
{\em Supersymmetry and supergravity} 
(Princeton Univ. Press, Princeton, 1983).

\bibitem{KLYT}
A. Klemm, W. Lerche, S. Yankielowicz and S. Theisen, 
Phys. Lett. B {\bf 344}, 169 (1995). 

\bibitem{KLT}
A. Klemm, W. Lerche and S. Theisen, 
Int. J. Mod. Phys. A {\bf 11}, 1929 (1996). 

\bibitem{APS}
P. C. Argyres and A. Faraggi, 
Phys. Rev. Lett. {\bf 74}, 3931 (1995). 

\bibitem{HO}
A. Hanany and Y. Oz, 
Nucl. Phys. B {\bf 452}, 283 (1995). 

\bibitem{HST} 
P. S. Howe, K. S. Stelle and P. K. Townsend, 
Nucl. Phys. B {\bf 214}, 519 (1983); 
Nucl. Phys. B {\bf 236}, 125 (1984). 

\bibitem{HSW}
P. S. Howe, K. S. Stelle and P. C. West, 
Phys. Lett. B {\bf 124}, 55 (1983). 

\bibitem{Sei}
N. Seiberg, 
Phys. Lett. B {\bf 206}, 75 (1988). 

\bibitem{AAG}
M. R. Abolhasani, M. Alishahiha and A. M. Ghezelbash, 
Nucl. Phys. B {\bf 480}, 279 (1996). 

\bibitem{AS}
P. C. Argyres, and A. D. Shapere, 
Nucl. Phys. B {\bf 461}, 437 (1996). 

\bibitem{DS1}
U. H. Danielsson and B. Sundborg, 
Phys. Lett. B {\bf 358}, 273 (1995). 

\bibitem{Han}
A. Hanany, 
Nucl. Phys. B {\bf 466}, 85 (1996). 

\bibitem{BL}
A. Brandhuber and K. Landsteiner, 
Phys. Lett. B {\bf 358}, 73 (1995). 

\bibitem{IS}
K. Ito, and N. Sasakura, 
Nucl. Phys. B {\bf 484}, 141 (1997). 

\bibitem{AF}
P. C. Argyres, M. R. Plesser and A. D. Shapere, 
Phys. Rev. Lett. {\bf 75}, 1699 (1995). 

\bibitem{DS2}
U. H. Danielsson and B. Sundborg, 
Phys. Lett. B {\bf 370}, 83 (1996). 

\bibitem{AAM}
M. Alishahiha, F. Ardalan and F. Mansouri, 
Phys. Lett. B {\bf 381}, 446 (1996).  

\bibitem{LPG}
K. Landsteiner, J. M. Pierre and S. B. Giddings, 
Phys. Rev. D {\bf 55}, 2367 (1997). 

\bibitem{Gray}
J. J. Gray, 
Bull. (New Ser.) Amer. Math. Soc. {\bf 10}, No.1, 
1 (1984).  

\bibitem{Ali1}
M. Alishahiha, 
Phys. Lett. B {\bf 398}, 100 (1997); Phys. Lett. B {\bf 418}, 317 (1998). 

\bibitem{KTS}
A. Klemm, S. Theisen and M. G. Schmidt, 
Int. J. Mod. Phys. A {\bf 7}, 6215 (1992).

\bibitem{IMNS1}
J. M. Isidro, A. Mukherjee, J. P. Nunes and H. J. Schnitzer, 
Nucl. Phys. B {\bf 492}, 647 (1997); 
Nucl. Phys. B {\bf 502}, 363 (1997).

\bibitem{SK}
H. M. Srivastava and P. W. Karlsson, 
{\em Multiple Gaussian hypergeometric series} 
(Ellis Horwood Limited, West Sussex, 1985).

\bibitem{S}
L. J. Slater, 
{\em Generalized hypergeometric functions} 
(Cambridge University Press, Cambridge, 1966).

\bibitem{E}
A. Erd\'{e}lyi, 
{\em Higher transcendental functions} 
(McGraw-Hill, New York, 1966). 

\bibitem{Exton}
H. Exton, 
{\em Multiple hypergeometric functions and applications}, 
(Ellis Horwood Limited, Chichester, 1976).

\bibitem{Kim}
T. Kimura, 
{\em Hypergeometric functions of two variables}, 
Seminar note in mathematics, Univ. of Tokyo, 1973. 

\bibitem{MSS}
T. Masuda, T. Sasaki and H. Suzuki, 
Int. J. Mod. Phys. A {\bf 13}, 3121 (1998). 

\bibitem{IY1}
K. Ito and S.-K. Yang, 
Phys. Lett. B {\bf 366}, 165 (1996). 

\bibitem{Ryang}
S. Ryang, 
Phys. Lett. B {\bf 365}, 113 (1996).

\bibitem{Ohta}
Y. Ohta, 
J. Math. Phys. {\bf 37}, 6074 (1996); J. Math. Phys. {\bf 38}, 682 (1997).

\bibitem{FP}
D. Finnell and P. Pouliot, 
Nucl. Phys. B {\bf 453}, 225 (1995).

\bibitem{IS3}
K. Ito and N. Sasakura,
Phys. Lett. B {\bf 382}, 95 (1996); 
Mod. Phys. Lett. A {\bf 12}, 205 (1997).  

\bibitem{Slat}
M. J. Slater, 
Phys. Lett. B {\bf 403}, 57 (1997).

\bibitem{DKM1}
N. Dorey, V. V. Khoze and M. P. Mattis,  
Phys. Lett. B {\bf 388}, 324 (1996); 
Phys. Rev. D {\bf 54}, 2921 (1996).  

\bibitem{DKM3}
N. Dorey, V. V. Khoze and M. P. Mattis,  
Phys. Rev. D {\bf 54}, 7832 (1996).

\bibitem{AHSW}
H. Aoyama, T. Harano, M. Sato and S. Wada, 
Phys. Lett. B {\bf 388}, 331 (1996).

\bibitem{HS}
T. Harano and M. Sato, 
Nucl. Phys. B {\bf 484}, 167 (1997).  

\bibitem{EF2}
H. Ewen and K. F\"{o}rger, 
Int. J. Mod. Phys. A {\bf 12}, 4725 (1997).

\bibitem{MW}
E. J. Martinec and N. P. Warner, 
Nucl. Phys. B {\bf 459}, 97 (1996). 

\bibitem{LW}
W. Lerche and N. P. Warner, 
Phys. Lett. B {\bf 423}, 79 (1998).

\bibitem{EY2}
T. Eguchi and S.-K. Yang, 
Phys. Lett. B{\bf 394}, 315 (1997).

\bibitem{Ito}
K. Ito, 
Phys. Lett. B {\bf 406}, 54 (1997). 

\bibitem{Ghe}
A. M. Ghezelbash, 
Phys. Lett. B {\bf 423}, 87 (1998). 

\bibitem{MS1}
T. Masuda and H. Suzuki, 
Int. J. Mod. Phys. A {\bf 13}, 1495 (1998); 
Int. J. Mod. Phys. A {\bf 12}, 3413 (1997).

\bibitem{DKP1}
E. D'Hoker, Krichever L. M. and D. H. Phong, 
Nucl. Phys. B {\bf 489}, 179 (1997); 
Nucl. Phys. B {\bf 489}, 211 (1997).  

\bibitem{DP}
E. D'Hoker and D. H. Phong, 
Phys. Lett. B {\bf 397}, 94 (1997). 

\bibitem{Mat}
M. Matone, 
Phys. Lett. B {\bf 357}, 342 (1995).  

\bibitem{FT}
F. Fucito and G. Travaglini, 
Phys. Rev. D {\bf 55}, 1099 (1997).

\bibitem{STY}
J. Sonnenschein, S. Theisen and S. Yankielowicz, 
Phys. Lett. B {\bf 367}, 145 (1996). 

\bibitem{EY}
T. Eguchi and S.-K. Yang, 
Mod. Phys. Lett. A {\bf 11}, 131 (1996). 

\bibitem{HW}
P. S. Howe and P. C. West, 
Nucl. Phys. B {\bf 486}, 425 (1997).

\bibitem{KO}
H. Kanno and Y. Ohta, 
Nucl. Phys. B {\bf 530}, 73 (1998). 

\bibitem{Ohta3}
Y. Ohta, 
Differential equations for scaling relation in 
$N=2$ supersymmetric SU(2) Yang-Mills theory 
coupled with massive hypermultiplet, to be published in J. Math. Phys, 
hep-th/9809180. 

\bibitem{EF1}
H. Ewen, K. F\"{o}rger and S. Theisen, 
Nucl. Phys. B. {\bf 485}, 63 (1997).  

\bibitem{Dub}
B. Dubrovin, 
Geometry of 2d topological field theories, hep-th/9407018. 

\bibitem{IY3}
K. Ito and S.-K. Yang, 
Phys. Lett. B {\bf 415}, 45 (1997); 
Phys. Lett. B {\bf 433}, 56 (1998).

\bibitem{MMM1}
A. Marshakov, A. Mironov and A. Morozov, 
Phys. Lett. B {\bf 389}, 43 (1996); 
Mod. Phys. Lett. A {\bf 12}, 773 (1997).

\bibitem{MMM3}
A. Marshakov, A. Mironov and A. Morozov, 
More evidence for the WDVV equations in $N=2$ 
SUSY Yang-Mills theories, hep-th/9701123.

\bibitem{IXY}
K. Ito, C.-S. Xiong and S.-K. Yang, 
Phys. Lett. B {\bf 441}, 155 (1998).

\bibitem{Ohta2}
Y. Ohta, 
Picard-Fuchs ordinary differential systems in 
$N=2$ supersymmetric Yang-Mills theories, to be published in 
J. Math. Phys, hep-th/9812085.

\bibitem{Nek}
N. Nekrasov, 
Nucl. Phys. B {\bf 531}, 323 (1998).


\end{thebibliography}
\end{document}